# Cloud Computing Through Mobile-Learning

N.Mallikharjuna Rao[1]
Associate Professor
Annamacharya P.G College of Computer Studies,
Rajampet, AP, India
e-mail:drmallik2009@gmail.com

C.Sasidhar[2]
Assistant Professor
Annamacharya P.G College of Computer Studies,
Rajampet, AP, India
e-mail: sasicmca39@gmail.com

V. Satyendra Kumar[3]
Assistant Professor
Annamacharya Institute of Technology and Sciences,
Rajampet, AP, India
e-mail: sati2all@gmail.com

*Abstract-* **Cloud computing is the new technology that has various advantages and it is an adoptable technology in this present scenario. The main advantage of the cloud computing is that this technology reduces the cost effectiveness for the implementation of the Hardware, software and License for all. This is the better peak time to analyze the cloud and its implementation and better use it for the development of the quality and low cost education for all over the world. In this paper, we discuss how to influence on cloud computing and influence on this technology to take education to a wider mass of students over the country. We believe cloud computing will surely improve the current system of education and improve quality at an affordable cost.**

## I. INTRODUCTION

Cloud Computing has been one of the most booming technology among the professional of Information Technology and also the Business due to its Elasticity in the space occupation and also the better support for the software and the Infrastructure it attracts more technology specialist towards it. Cloud plays the vital role in the Smart Economy, and the possible regulatory changes required in implementing better Applications by using the potential of Cloud Computing [1][2][3].

The main advantage of the cloud is that it gives the low cost implementation for infrastructure and some higher business units like Google, IBM, and Microsoft offer the cloud for Free of cost for the Education system, so it can be used in right way which will provide high quality education. In this paper, we discussed back ground of this paper in section 2, section 3 presented present scenario for existing systems, in section 4 discussed the proposed cloud system for education and section 5 illustrated merits and section 6 concluded with advantages.

## II BACKGROUND

The term *cloud computing* is being bandied about a lot these days, mainly in the context of the future of the web. But cloud computing potential doesn't begin and end with the personal computer's transformation into a thin client - the mobile platform is going to be heavily impacted by this technology as well. At least that's the analysis being put forth by ABI Research. Their recent report, Mobile Cloud Computing, theorizes that the cloud will soon become a disruptive force in the mobile world, eventually becoming the dominant way in which mobile applications operate.

With a Western-centric view of the world, it can sometimes be hard to remember that not everyone owns a smart phone. There are still a large number of markets worldwide where the dominant phone is a feature phone. While it's true that smart phones will grow in percentage and feature phones will become more sophisticated in time, these lower-end phones are not going away anytime soon. And it's their very existence which will help drive the mobile cloud computing trend.

Not only is there a broader audience using feature phones in the world, there are also more web developers capable of building mobile web applications than there are developers for any other type of mobile device. Those factors, combined with the fact that feature phones themselves are becoming more capable with smarter built-in web browsers and more alternative browsers available for download, will have an impact on mobile cloud computing growth. As per the above statements, we are proposing to use any software applications on mobiles. They can use in even villages of rural areas in India, because, in India most of the part is covered in rural areas.

## III. PRESENT SCENARIO

In this scenario cloud computing is being looked upon by experts in various domains because of its advantages. Cloud has been used in the business oriented unit and in the current education system in India the teaching via web is not so widely available and adapted. Even if it is available, it is provided at a very high cost. This is mainly because of the high cost of data storage and the software they make use of. Cloud has generated many resources which can be used by various educational institutions and streams where their existing/proposed web based learning systems can be implemented at low cost.

### A. Benefits of Cloud Computing

The advantages that come with cloud computing can help resolving some of the common challenges one might have while supporting an educational institution. [4][5].

*1) Cost*
One can choose a subscription or in, some cases, pay-as-you-go plan –whichever works best with that organization business model.

*2) Flexibility*
Infrastructure can be scaled to maximize investments. Cloud computing allows dynamic scalability as demands fluctuate.

*3) Accessibility*
This help makes data and services publicly available without make vulnerable sensitive information.

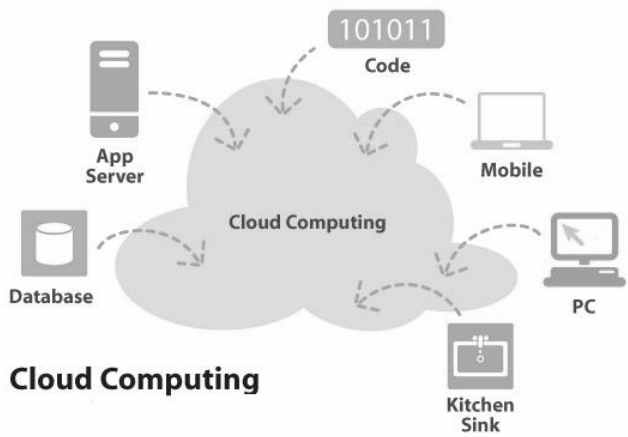

Figure 1: Cloud computing with various components

Some would resort to a cloud computing vendor because of the lack of resources while others have the resources to build their cloud computing applications, platforms and hardware. But either way, components have to be implemented with the expectation of optimal performance when we are using through mobile terminals [7].

*4) The Client – The End User*

Everything ends with the client (mobile). The hardware components, the application and everything else developed for cloud computing will be used in the client. Without the client, nothing will be possible. The client could come in two forms: the hardware component or the combination of software and hardware components. Although it's a common conception that cloud computing solely relies on the cloud (internet), there are certain systems that requires pre-installed applications to ensure smooth transition. In this work, all the pre-installed applications can view by mobile devices though clouds. The hardware on the other hand will be the platform where everything has to be launched. Optimization is based on two fronts: the local hardware capacity and the software security. Through optimized hardware with security, the application will launch seamlessly with mobile devices [7].

Cloud computing always has a purpose. One of the main reasons cloud computing become popular is due to the adoption of businesses as the easier way to implement business processes. Cloud computing is all about processes and the services launched through mobile cloud computing always has to deal with processes with an expected output.

*B. Services in Cloud Computing*

Infrastructure as a Service**.** One can get on-demand computing and storage to host, scale, and manage applications and services. Using Microsoft data centers means one can scale with ease and speed to meet the infrastructure needs of that entire organization or individual departments within it, globally or locally [6].

Platform as a Service. The windows azure cloud platform as a service consists of an operating system, a fully relational database, message-based service bus, and a claims access controller providing security-enhanced connectivity and federated access for on premise applications. As a family of on-demand services, the Windows Azure platforms offers organization a familiar development experience, on-demand scalability, and reduce time to market the applications.

Software as a Service. Microsoft hosts online services that provide faculty, staff, and students with a consistent experience across multiple devices.

Microsoft Live at edu provides students, staff, faculty, and alumni long-term, primary e-mail addresses and other applications that they can use to collaborate and communicate online— all at no cost to your education institution.

Exchange Hosted Services offers online tools to help organizations protect themselves from spam and malware, satisfy retention requirements for e-discovery and compliance, encrypt data to preserve confidentiality, and maintain access to e-mail during and after emergency situations.

Microsoft Dynamics CRM Online provides management solutions deployed through Microsoft Office Outlook or an Internet browser to help customers efficiently automate workflows and centralize information. Office Web Apps provide on-demand access to the Web-based version of the Microsoft Office suite of applications, including Office Word, Office Excel, and Office PowerPoint.'

*C. Cloud computing usage*

The cloud plays the main role in the business role and also it is the only elastic data centre which wrapped around various new technologies into it. The technology is most probably used in the business oriented scenario than the service motivated organization as per the survey did by us. According to the Survey made during the month of October 2010 based on the questionnaire prepared by us we found that a major part of the survey group knew about cloud computing (Figure. 1), 69% knew that cloud is used in business, 12% knew it is used in education, 88% agree to implement the cloud for education sector, 94% believes that the cloud technology can reduce the cost of high quality education system and most of them are unaware that the cloud is also offered at low cost.

The following chart describes the statistical report generated based on various surveys done on the cloud computing methodologies and also based on the usage of the cloud in various sectors like data sharing, web mail services, store personal photos online applications such as Google Documents or Adobe Photoshop Express, store personal videos online, pay to store computer files online and also for back up hard drive to an online site.

*D. Requirements for Cloud*

In the previous generation of the information technology the data sharing which led the path for the knowledge sharing was not used by the users globally, in this generation the various streams have the knowledge of e-Learning and the Mobile based learning. In this present context the usage of the central data centre is a easy process for the education system however the cost of implementation and the maintenance of the data storage space and also the load capability also software licensing depends on the real time usage of these systems. Business streams can make revenue out of those expenses whereas for educational institutions which really want to motivate the learners and want to offer a quality education at affordable cost can achieve this by spending a large amount. This can be overcome by the present cloud computing technology that is "Pay as Use" (PAU).

## IV PROPOSED SYSTEM FOR EDUCATION

In this proposed system the main advantage is that the cloud computing has been used here to overcome all the drawbacks of the present system. Cloud computing is designed as that it can support any group of the users in all the criteria like the software, hardware, infrastructure, storage space everything is provided by the cloud and also the user has to pay as much as they need and don't want to pay for unused and not required as we do for the present data centers.

This cloud system has the cloud model and the client model for its implementation. In the cloud model it is designed as if its suits all the requirements and also the basic consideration. Here the various devices are used and also it has the device control, memory control, load control and several organizing units with the resources like networks, high bandwidth support also the main Security constrains with the filters and the firewall for the better maintenance of the data with proper backup methodology and now the system is ready to provide the data from the cloud.

The main process of this cloud model (Figure. 2) is that it provides the data manipulation operation with the load control and also high authentication and the authorization that too based on the external needs the size of the cloud usage varies so flexible and elastic in means of the data storage and the load accessibility but no compromise on security or the data backup.

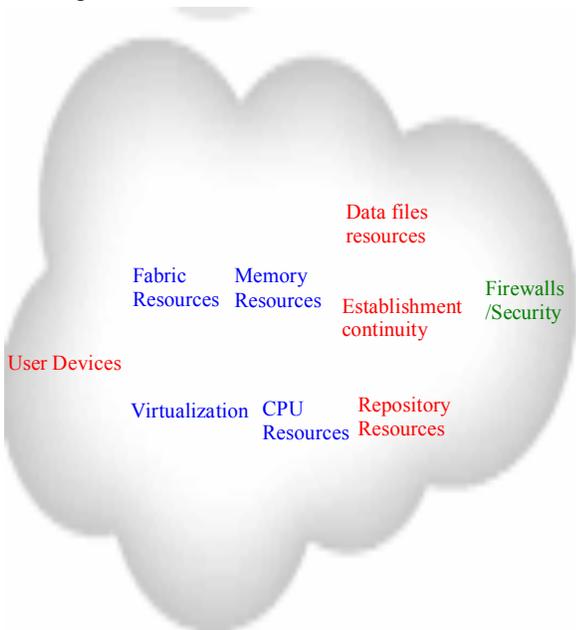

Figure 2: The Cloud Model Infrastructure and its Integrated resources

In this client model (Figure. 3) where the data is to be distributed, so that knowledge resources will be used by the all sorts of user in the education streams. Here the applications are developed which will mainly concentrate on the invalid usage of cloud and the data has to be managed and send to the user based on the various data centers available so this client model will check for the registration and valid clients to login into the system and use the application and also the security is maintained in this model so that the data are safeguarded. This model will be used to access the data and share the knowledge.

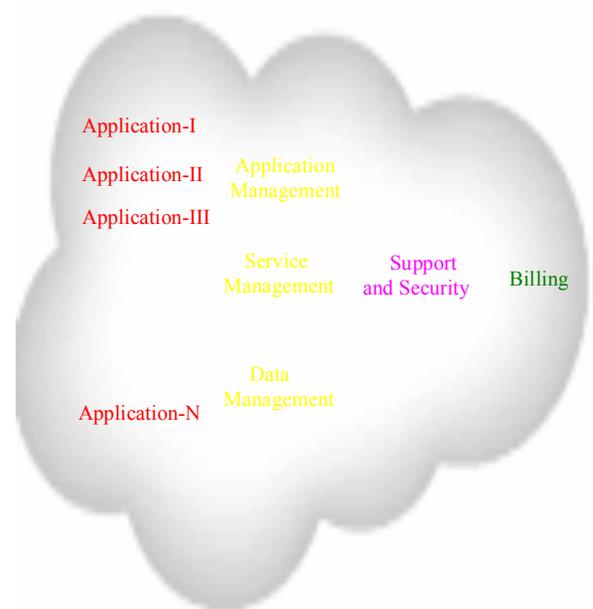

Figure 3: The Client Model Infrastructure and it's integrate resources and also the various processing layers in the model

### A. Mobile- Learning

This is a system which is implemented for education using cloud computing. The main objective of Mobile-Learning is that the learners can get the knowledge from the centralized shared resources at any time and any where they like to read that too at free of cost.

Mobile-learning is a system where one can learn through any source on topics of his choice without the need of storing everything in his device. As-you-pay and that much can you can use the services from the cloud data centers for learning selected topics over mobile phone even you in a small village or remote area. For example, if student want learn a JAVA technologies from his agricultural land and works.

### B. Functionality of Mobile-Learning

The person who wants to make use of Mobile-learning (Figure. 4) has to register and get the credentials to use it via web. It can also be downloading as a mobile application which will be installed in the mobile and through the GPRS/WIFI connectivity they can access the content over the cloud and the user can select among various available topics the one he needs. The topic might be Text based documents, audio and video files which will be buffered from the cloud to that mobile user and downloaded in the mobile if the memory is available in the mobile(if the user wishes to do so). The user can read the documents, look at the video tutorials, listen to lectures or seminars and finally they can take up self assessments. They will be given a results analysis so that they can evaluate their strengths & weaknesses on their own. This system helps to "Learn while you roam" and also education for all at any time any where globally. Experts can also share their valid tutorials in to the cloud for development of the education community.

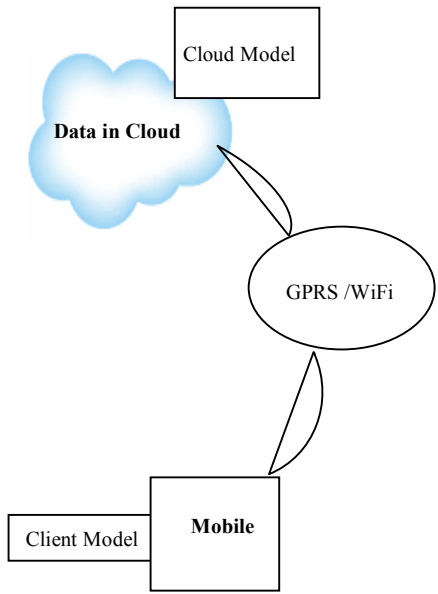

Figure 4: The Process flow of Mobile-Learning Cloud Computing

*C. Mobile-Learning Cloud Model*

In this cloud model (Figure 5) the User will access the cloud space using his/her credentials so that the required data will be shared from the cloud based on the client request only for the authenticated user.

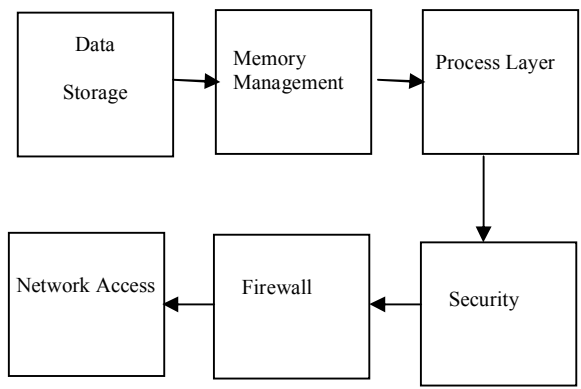

Figure 5: The Process flow of Mobile-Learning Cloud Model

In this paper, the process flow of mobile learning cloud as we shown in figure 5 which is having 6 steps. Data storage is used for storing the huge data where users are retrieving or handle the data from the data centers. Memory management is organizing and manages the data which is coming from clouds to mobile subscribers and process layer is interacting with security firewalls and memory management.

*D. Mobile-Learning Client Model*

In this client model (Figure 6) the user has to download this application and install in their Personal Digital Assistance (PDA) devices or in their mobile phones. The user has to connect to GPRS / Bluetooth / Wi-Fi and connect to the cloud network and get the required topics and based on the selected topic the materials will be downloaded to the mobile for the reading process.

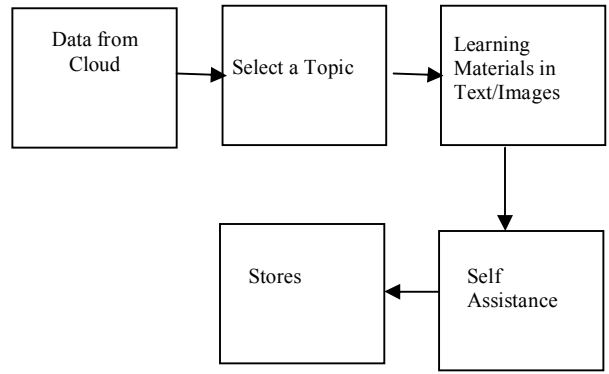

Figure 6: The Process flow of Mobile-Learning Client Model

As we shown in figure 6, the mobile users retrieve the data either in the form of text/video/voice from the cloud center. The subscribers are select which they want to download or retrieve from the data centers with the help of self assistance and keep downloaded data at mobile database.

*E. Advantages with Cloud in Mobile-Learning*

In this Mobile-Learning the cloud plays a vital role because the data sharing is the very important role of this learning system, so cloud takes the responsibility of data sharing security and also the load management during the peak hours of access without affecting the network band access. The cloud helps to increase the storage space if the data content are posted more by the users and also during peak hours the total number of user who uses the system will be increased so the load has to be tolerated automatically. Some of the companies offer the cloud at free of cost or at economy prices so this cloud computing will helps in offering the very quality high class education at affordable price.

Today there are more direct applications for teaching and learning as opposed to simple platform-independent tools and scalable data storage

The web search found that organizations of all sizes were using mobile devices for learning because technological advances meant that there was no longer the need for large infrastructure and support costs, and even small enterprises could deliver mobile learning simply by structuring learning around web-based content that could be accessed from web-enabled mobile devices.

The economics of cloud computing provide a compelling argument for mobile learning. Cloud-based applications can provide students and teachers with free or low-cost alternatives to expensive, proprietary productivity tools. For many institutions, cloud computing offers a cost-effective solution to the problem of how to provide services, data storage, and computing power to a growing number of Internet users without investing capital in physical machines that need to be maintained and upgraded on-site.

Teachers don't have to worry about using outdated or different versions of software. As an Algebra department, we often use Classroom Performance System (CPS) and when we send our test files to each other, this often generates errors because we don't have the same version installed. The same goes with Microsoft Word documents. The cloud will take care of issues like these.

Students and teachers will have 24/7 access to not only their files, but their applications as well (provided they have Internet access). This means that if we need to create an assessment using the CPS software, we don't have to worry about having it installed on every single computer we need to access it from.

The Cloud Computing Opportunity by the Numbers, a multitude of interesting and convincing figures that back the claim that there is a huge opportunity arising in the cloud. With all due credit for the stats going to the article from Reuven's website: ElasticVapor, we wanted to share some of these interesting points:

- There are 50 Million servers worldwide today. By 2013 60% of server workload will be virtualized
- In 2008 the amount of digital information increased by 73%
- There were 360,985,492 internet users in 2000. In 2009 that number increased to 1,802,330,457. That's roughly 27% of the entire world population.
- 50% of the servers sold worldwide are destined for use in a data centre (the average data centre uses 20 megawatts, 10 times more than data centers in 2000 used.)
- Merrill Lynch predicts that the cloud computing market will reach $ 160 billion by 2011.
- IBM claims Cloud cuts IT labor costs by up to 50% and improves capital utilization by 75%.

V MERITS OF CLOUD COMPUTING MOBILE-LEARNING

We described so many advantages offered by cloud computing in mobile education. Following are the some of important merits with mobile computing.

A. Lower costs.

You don't need a high-powered and high-priced computer to run cloud computing web-based applications, since applications run in the cloud, not on the desktop PC. In a Simple way we can run such high configured applications on your mobile with cheap cost. When you're using web-based applications on mobiles need not required any memory space and as no software programs have to be loaded and no document files need to be saved.

B. Improved performance

With fewer overfed programs hogging your mobile memory, we will see better performance from your mobile device. Put it simply, mobiles in a cloud computing system boot and run faster because they have fewer programs and processes loaded into mobile memory.

C. Reduced software costs.

Instead of purchasing expensive software applications, you can get most of what you need for free or in least prices on mobile devices even in rural area.

D. Instant software updates

Another software-related advantage in cloud computing is that we are no longer faced with choosing between obsolete software and high upgrade costs. When the app is web-based, updates happen automatically and are available the next time you log on to the cloud. When you access a web-based application, you get the latest version without needing to pay for or download an upgrade with our mobile device.

E. Improved document format compatibility

In mobile cloud computing, we have more compatibility for opening the files, applications easily with installation of several software's on mobile device.

F. Increased data reliability

In desktop computing, in which a hard disk crash can destroy all your valuable data, a computer crashing in the cloud should not affect the storage of your data. Even, if your personal computer crashes, all your data is still out there in the cloud, still accessible. Hence, cloud computing is the ultimate in data-safe computing.

G. Universal document access

All your documents are instantly available from wherever you are and there is simply no need to take your documents with you.

H. Device independence.

Finally, here's the ultimate cloud computing advantage: You're no longer tethered to a single computer or network. Change computers, and your existing applications and documents follow you through the cloud. Move to a portable device which is mobile phone, and your applications and documents are still available. There is no need to buy a special version of a program for a particular device, or to save your document in a device-specific format.

## VI CONCLUSIONS

The cloud computing has the significant scope to change the whole education system. In present scenario the e-learning is getting the popularity and this application in cloud computing will surely help in the development of the education offered to poor people which will increase the quality of education offered to them. Cloud based education will help the students, staff, Trainers, Institutions and also the learners to a very high extent and mainly students from rural parts of the world will get an opportunity to get the knowledge shared by the professor on other part of the world. Even governments can take initiatives to implement this system in schools and colleges in future and we believe that this will happen soon.

## AUTHORS PROFILE

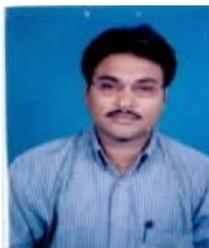

N.Mallikharjuna Rao is presently working as Associate Professor in the department of Master of Computer Applications at Annamacharya PG college of Computer Studies, Rajampet and having more than 12 years of Experience in Teaching UG and PG courses. He received his B.Sc Computer Science from Andhra University in 1995, Master of Computer Applications (MCA) from Acharya Nagarjuna University in 1998, Master of Philosophy in Computer science from Madurai Kamarj University, Tamilnadu, in India and Master of Technology in Computer Science and Engineering from Allahabad Agricultural University, India. He is a life Member in ISTE and Member in IEEE, IACSIT. He is a research scholar in Acharya Nagarjuna University under the esteemed guidance of Dr.M.M.Naidu, Principal, SV.Univeristy, Tirupathi, and AP.

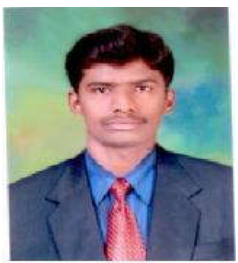

Mr.V.Sathyendra Kumar is working as Assistant Professor in the Department of MCA at Annamacharya Institute of Technology & Sciences, Rajampet. He has more than 4 years of Teaching Experience for PG. He is life member in ISTE.

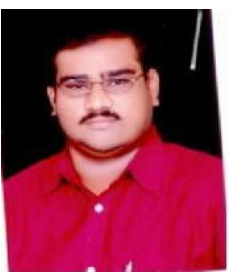

Mr. C. Sasidhar is working as Assistant Professor in the Department of MCA, Annamacharya P.G College of Computer Studies, Rajampet. He has more than 5 years of Teaching Experience for PG Courses. He received his MCA and M.Tech (CS) from JNTUA in the year 2005, 2010 respectively. He is life member in ISTE.